\documentclass[aps,pre,floatfix,twocolumn,superscriptaddress]{revtex4-2}

\usepackage{amsmath}
\usepackage{amsfonts}
\usepackage{amssymb}
\usepackage{graphicx}
\usepackage{parskip}

\begin{document}

\title{Localization of quantum walk with classical randomness: Comparison between manual methods and supervised machine learning}

\author{Christopher Mastandrea}
\affiliation{Department of Physics, University of California, Merced, CA 95343, USA}

\author{Chih-Chun Chien}
\email{cchien5@ucmerced.edu}
\affiliation{Department of Physics, University of California, Merced, CA 95343, USA}

\begin{abstract}
A transition of quantum walk induced by classical randomness changes the probability distribution of the walker from a two-peak structure to a single-peak one when the random parameter exceeds a critical value. We first establish the generality of the localization by showing its emergence in the presence of random rotation or translation. The transition point can be located manually by examining the probability distribution, momentum of inertia, and inverse participation ratio. As a comparison, we implement three supervised machine learning methods, the support vector machine (SVM), multi-layer perceptron neural network, and convolutional neural network with the same data and show they are able to identify the transition. While the SVM sometimes underestimate the exponents compared to the manual methods, the two neural-network methods show more deviation for the case with random translation due to the fluctuating probability distributions. Our work illustrates potentials and challenges facing machine learning of physical systems with mixed quantum and classical probabilities. 
\end{abstract}

\maketitle

\section{Introduction}
While classical random walk finds broad applications in physics, chemistry, biology, finance, and many other places~\cite{PhysRevLett.119.230601, Biomed.1471.2105.10.17, rsif.2008.0014}, the simplest quantum analogue, the quantum walk (QW), exhibits interesting probabilistic behavior due to the underlying wavefunction even if all the operations are deterministic~\cite{PhysRevA.97.012308}. Recently, there is a trend in studying various quantum walks with classical randomness~\cite{PhysRevA.77.022302,PhysRevA.83.022320,PhysRevLett.106.180403,crespi2013anderson,PhysRevA.92.052311}, in the sense that the parameter set or geometry of QW is drawn from a classical probability distribution. The resulting probability distribution thus has contributions from both classical and quantum randomness. As the classical randomness increases, a transition from delocalization to localization emerges, which has stimulated lasting research interest.

There have been experimental demonstrations of discrete-time quantum walk with classical randomness in the quantum operators of the evolution and exhibitions of the transition from quantum dynamics with multi-peak probability distributions in real or momentum space to classical-like dynamics with Gaussian-like probability distributions. For example,
phase-disordered photons ~\cite{PhysRevLett.106.180403,PhysRevLett.104.153602}, 
trapped ions with randomized phases~\cite{PhysRevLett.104.100503}, superconducting qubits with random frequencies~\cite{PhysRevA.89.022309}, and neutral atoms with random microwave pulses~\cite{PhysRevLett.121.070402} have been implemented to demonstrate the transition induced by classical randomness.

Previous theoretical investigations~\cite{1DFiberLoopALocal,DTQWRandomLattice,PhysRevA.92.052311,Yao23} have suggested possible resemblance between the localization transition in quantum walk with classical randomness and the Anderson localization~\cite{PhysRev.109.1492}, a paradigmatic phenomenon in condensed matter physics. 
The Anderson localization was first developed in electronic transport, where  electrons in a solid encounter a localization transition as the parameters from the underlying materials are drawn from a classical probability distribution. 
Ref.~\cite{Abrahamsscaling} proposed a scaling theory applicable to higher dimensions and showed that in one and two dimensions, any amount of disorder will lead to electron localization in the thermodynamic limit. Nevertheless, in a finite system, the Anderson localization only occurs when the system size exceeds the localization length. For example, Ref. \cite{1DFiberLoopALocal} evaluates the localization length inferred from the inverse participation ratio. 
Although the Anderson localization typically has spatial randomness while QW usually has temporal randomness, they both exhibit localization transitions due to classical randomness.
As our simulation results will show, for quantum walk in finite-size systems, the classical randomness has to exceed a threshold for the system to be localized. Therefore, even if imperfections or fluctuations in the apparatus are unavoidable and may introduce classical randomness into quantum-walk experiments, the quantum behavior may still survive when the threshold is not crossed. 
Other types of QW, such as Grover QW~\cite{PhysRevA.69.052323,Zeng17} and QW in  inhomogeneous~\cite{PhysRevE.82.031122,PhysRevA.85.012329} or driven Floquet systems~\cite{PhysRevB.96.144204}, may also exhibit localization in theory.

Meanwhile, machine learning has become a powerful tool in physics problems like phase transitions~\cite{Carrasquilla2017,PhysRevB.102.134213}, dynamics~\cite{PhysRevLett.120.225502}, gravitational wave detection ~\cite{PhysRevD.102.063015,PhysRevD.105.083007}, searching for new physics~\cite{Karagiorgi22}, etc. More applications are reviewed in Refs.~\cite{Tanaka19,RevModPhys.91.045002,MEHTA20191,Karniadakis2021,Bedolla21}. Since the identification of the localization transition of QW with classical randomness is a demanding task when performed manually, we seek help from machine learning to automate the analysis. We test two elementary supervised machine learning methods, the support vector machine (SVM) and multi-layer perceptron neural network (MLP NN), and a more sophisticated one, the convolutional neural network (CNN). Those methods have been applied to universal classifications of classical jammed systems \cite{PhysRevLett.123.160602} and topological properties of materials~\cite{PhysRevB.101.245117}. Interestingly, we will show that depending on the type of randomness, the three methods may systematically underestimate the scaling behavior of the localization transition due to the fluctuating patterns and continuous transforms of the probability distributions.

To analyze the scaling behavior of the localization of QW with classical randomness, we introduce three types of classical probabilities, one discrete and the other two continuous, into the rotation and translation operators of the QW. We first use physical quantities, including the patterns of probability distribution, moment of inertia (MoI), and inverse participation ratio (IPR), to manually locate the critical point. We then test the hypothesis that the labor- and time- demanding task of determining the scaling behavior of the localization can be handled automatically by supervised machine learning. 
We will show that the analyses by the machine learning methods can differentiate the localized and delocalized regimes and exhibit scaling behavior. For QW with discrete random rotation, the machine learning methods produce exponents close to those from the manual methods. However, for QW with continuous random rotation, the machine-learning methods tend to under-estimate the exponents. In the case of QW with random translation, the NN-based methods exhibit more significant deviation in determining the exponent of the localization, possibly due to the more complicated structures of the probability distributions in the transition regime. Therefore, QW with classical randomness provides an example with both quantum and classical probabilities that is challenging to available supervised machine learning methods.

The rest of the paper is organized as follows. Sec.~\ref{sec:Theory} briefly reviews discrete-time quantum walk and introduces three types of classical randomness through the rotation or translation operators. Sec.~\ref{sec:Localization} presents the localization transition and documents the three manual and three machine-learning methods for analyzing the transition.
Sec.~\ref{sec:Comparison} compares the results from the manual and machine-learning methods and shows where the exponents of the localization transition from machine-learning methods agree or disagree with the manual methods. We also mention the Anderson localization and discuss implications for experimental or theoretical research. Finally, Sec.~\ref{sec:Conclusion} concludes our work.

\section{Theoretical background}\label{sec:Theory}
\subsection{Discrete quantum walk}
We consider discrete-time QW on a 1D lattice following the description of Ref.~\cite{QWalkSpringer}, beginning with a quantum walker described by a wavefunction defined on a 1D lattice with two internal orthonormal coin states ($|+\rangle, |-\rangle$) on each site.
The total state of the walker is taken to be a superposition of the $|+\rangle, |-\rangle$ degrees of freedom, $| \psi \rangle = \sum_{x}(\alpha_{x,+}|x,+ \rangle + \alpha_{x,-}|x, -\rangle)$, where $x$ labels the lattice sites and $\alpha_{x,+} = \langle x, + | \psi \rangle$, $\alpha_{x,-} = \langle x, - | \psi \rangle$ are the complex-valued probability amplitudes for each coin state.

The QW uses two operators, the rotation operator $\mathcal{\hat{C}}$ and the translation operator $\mathcal{\hat{T}}$.
The translation operator is defined by  
$\mathcal{\hat{T}}|\psi_{x}, +\rangle = |\psi_{x+1}, +\rangle $ and
$\mathcal{\hat{T}}|\psi_{x}, -\rangle = |\psi_{x-1}, -\rangle $.
Thus, the translation operator acts to shift the $|+\rangle$ state by one lattice site in the positive direction and the $|-\rangle$ state by one site in the negative direction.
The rotation operator is defined as a general unitary operator acting on the coin states at each site and is given by
$\mathcal{\hat{C}} (\theta, \phi_{1}, \phi_{2}) = \begin{pmatrix} \cos(\theta) & e^{i\phi_{1}}\sin(\theta)  \\ e^{i\phi_{2}} \sin(\theta) & -e^{i(\phi_{1} + \phi_{2}})\cos(\theta) \\ \end{pmatrix}$.
The rotation operator mixes the $|+\rangle$ and $|-\rangle$ states at each lattice site, acting as the quantum analog to the coin flip of the classical random walk.
Throughout the paper, we will set $\phi_{1} = \phi_{2} = \frac{\pi}{2}$ so that the coin operator is only dependent on $\theta$, i.e, 
$\mathcal{\hat{C}} = \begin{pmatrix} \cos(\theta) & i\sin(\theta)  \\  i\sin(\theta) & \cos(\theta) \\ \end{pmatrix}$.\
The total state of the walker after some time steps $N$ is found through repeatedly applying the rotation and translation operators to the walker's initial state. $|\psi(t)\rangle = \mathcal{\hat{T}} \mathcal{\hat{C}} \mathcal{\hat{T}} \mathcal{\hat{C}} \cdots \mathcal{\hat{T}} \mathcal{\hat{C}} |\psi_{0} \rangle = (\mathcal{\hat{T}} \mathcal{\hat{C}})^{N} |\psi_{0} \rangle$.

If the quantum walk only runs up to a time $N$ in the simulation, the lattice considered has size of at least $(2 \times N) + 1$, so the boundary will not be exceeded if the walker starts in the middle. 
Here an extra lattice site to accommodate the initial location is added to the middle of the lattice, so that the lattice has a symmetric amount of steps in both directions. 
We consider an initial state with an equal superposition of the $|+\rangle$ and $|-\rangle$ states at the origin, $| \Psi_{0} \rangle = \frac{1}{\sqrt{2}} (|0,+ \rangle + | 0,- \rangle)$ and study how the wavefunction spreads out in the lattice.

\subsection{Quantum walk with classical randomness}
In addition to the quantum probability from the wavefunction in a quantum walk, we introduce classical randomness that may come from imperfections or fluctuations of the apparatus or environment. There are many ways to add classical randomness to quantum walk. As mentioned, Ref.~\cite{1DFiberLoopALocal} allows random values of the phase angle $\phi$ in the rotation operator. Here we introduce three other types of classical randomness that will be analyzed later.

\subsubsection{Discrete random angles in rotation}
We first introduce classical randomness to  the quantum walk by using two rotation operators instead of just one. The two rotation operators $\mathcal{\hat{C}}_{1,2}$ have their $\theta$ values given by $\theta_{1} = \theta_{0} + \Delta\theta$ and $\theta_{2} = \theta_{0} - \Delta\theta$ while all the $\phi$-angles are set to $\pi/2$. 
At each step of the walk, we flip a fair classical coin to choose from $\theta_{1,2}$. Explicitly, $P(\theta=\theta_1)=1/2=P(\theta=\theta_2)$ at each step.
The walk is evolved through the normal application of the chosen coin operator and the translation operator repeatedly. Therefore, the two rotation operators and the translation operator are deterministic, but the classical coin adds an additional probability distribution to the quantum walk.
In our simulations, we scan through values of $\Delta \theta$ over the range $0 \leq \Delta \theta \leq \theta_{0}$
to locate the localization transition.

\subsubsection{Continuous random angle in rotation}
To connect to systems with continuous classical randomness, we consider a quantum walk with random rotation, where the angle $\theta$ of the rotation operator is determined by a uniform distribution, in contrast to the binary distribution in the previous case. Explicitly, the angle $\theta$ in each step of one simulation is given by $\theta(t)=\theta_0+\Delta\theta(t)$, where $\Delta\theta(t)$ is drawn from a uniform distribution between $(0,\Delta\theta_M)$ for a given $\Delta\theta_M$. The $\phi$-angles are fixed at $\pi/2$.
Next, we scan through different values of $\Delta\theta_M$ by running individual simulations for each value of $\Delta\theta_M$ and obtain the probability distributions from the wavefunction. We remark that Ref.~\cite{Yao23} also implements continuous changes of the rotation to study the localization and entanglement between the rotation and translation degrees of freedom of QW.

\subsubsection{Random translation}
We consider another type of classical randomness in quantum walk. In line with the quantum walk operation, there is only one fixed rotation operator throughout the walk. However, an inverse-translation operator is introduced and is defined by 
$\mathcal{\hat{T}}^{-1}|\psi_{x}, +\rangle = |\psi_{x-1}, +\rangle $ and
$\mathcal{\hat{T}}^{-1}|\psi_{x}, -\rangle = |\psi_{x+1}, -\rangle $,
such that it reverses the action of the original translation operator, $\mathcal{\hat{T}}^{-1}\mathcal{\hat{T}}|\psi_{x}\rangle = |\psi_{x}\rangle$.
We introduce  quantum walk with random translation by adding classical probability at every step to choose whether the translation  $\mathcal{\hat{T}}$ or the inverse translation $\mathcal{\hat{T}}^{-1}$ is applied after the rotation operator. A probability $P_{r}$ is assigned to the inverse translation. Explicitly, $P(\mathcal{\hat{T}}^{-1}) = P_{r}$ and  $P(\mathcal{\hat{T}}) = 1- P_{r}$. If $P_r = 0$, it is the QW without classical randomness while the maximal randomness occurs when $P_r=0.5$. If $P_r > 0.5$, the result is symmetric with respect to that with $1-P_r$ because of the parity symmetry between $\mathcal{\hat{T}}$ and $\mathcal{\hat{T}}^{-1}$.

\section{Localization transition}\label{sec:Localization}
The classical randomness can be increased by increasing $\Delta\theta$, $\Delta\theta_M$, or $P_r$ in the aforementioned models, and a localization transition starts to emerge. In the following, we discuss how the localization transition can be identified via different methods.

\subsection{Manual methods}
\subsubsection{Final probability distribution}
\begin{figure}
    \centering
    \includegraphics[width=\columnwidth]{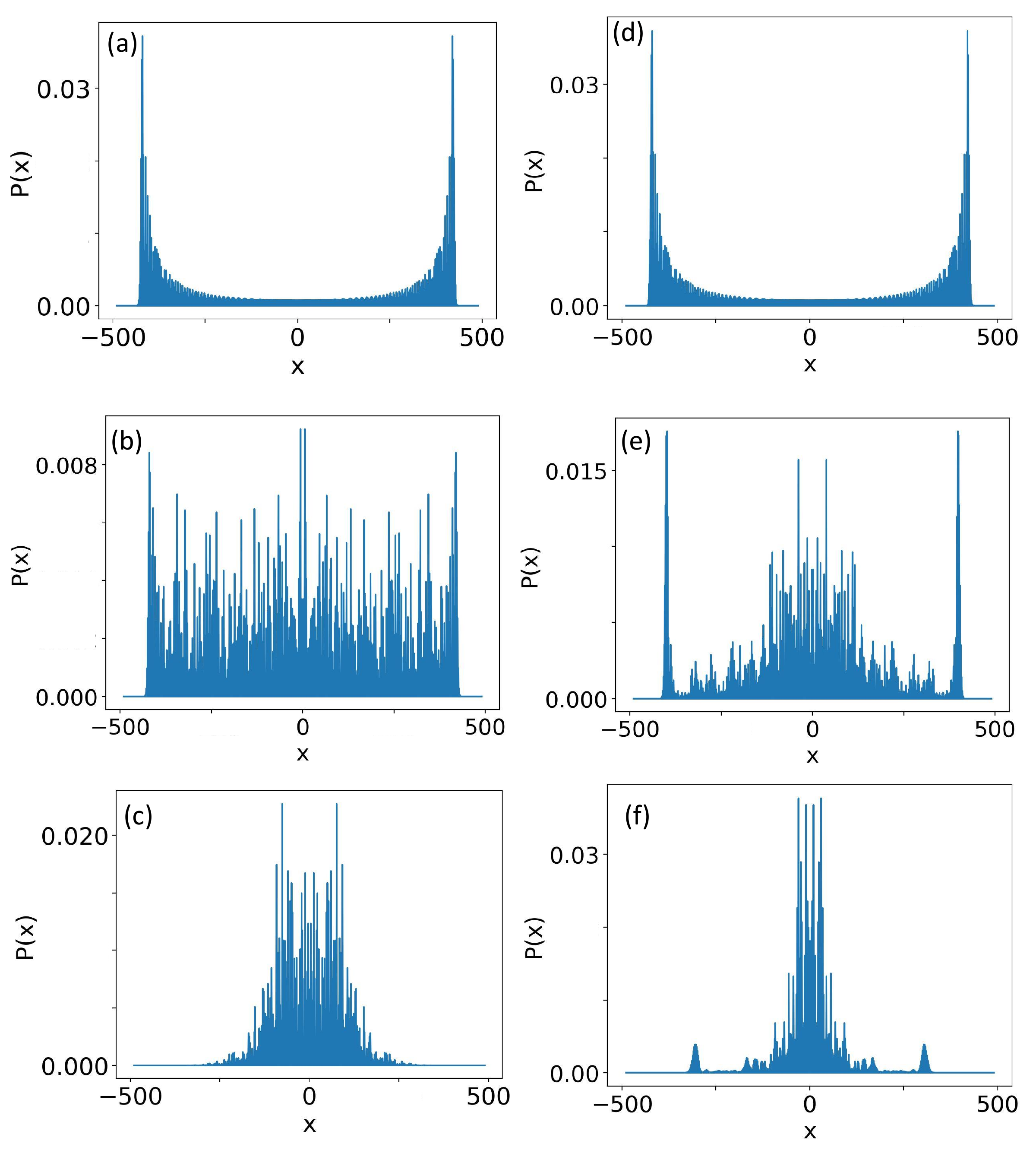}
    \caption{  Localization transition of QW with classical randomness: The probability distributions indicate the system in the delocalized (top row), critical (middle row), and localized (bottom row) regimes as the classical randomness increases. The left (right) column shows the case with discrete random rotation (random translation) with $\theta_{0} = \frac{\pi}{6}$ and $N = 490$. Panels (a), (b), (c) are at $\Delta\theta = 0$,  $\Delta\theta = 0.035$, and  $\Delta\theta = 0.09$, respectively. 
    Panels (d), (e), (f) are at $P_{r} = 0.005$, $P_{r} = 0.045$, and $P_{r} = 0.15$, respectively.
    \label{fig:ProbFigure}
    }
\end{figure}

The probability distribution of the walk at a given time can be found by summing over the probabilities on both coin states: $P(x) = \sum_{x}(|\psi_{+}(x)|^2 + |\psi_{-}(x)|^2)$.
Selective representatives of the final probability distributions of the cases with discrete random rotation and random translation are shown in Fig.~\ref{fig:ProbFigure}. One can see that when the classical randomness is weak, such as small $\Delta\theta$ or $P_r$, the quantum walker spreads out with two peaks in both directions of the lattice. However, when the classical randomness is strong, the probability distribution concentrates around the initial location with a single-peak structure, indicating localization of the walker. Therefore, a localization transition occurs at a critical value of the classical randomness. As shown in the middle panels of Fig.~\ref{fig:ProbFigure}, the probability distributions may behave differently in the transition regime for different types of classical randomness. For the discrete and continuous random rotations, the probability distribution becomes flat at the transition. However, the case with random translation exhibits a three-peak structure in the transition regime. 

By comparing the final probability distributions with fixed $N$ steps of evolution but different values of classical randomness, the transition region can be found visually by identifying the probability distributions that has a flattened-out pattern or a three-peak structure. Later on, we will see that the three-peak structure of QW with random translation imposes challenges for machine learning.

\subsubsection{Moment of inertia}
Next, we use the idea of moment of inertia (MoI) from classical mechanics to introduce another measure of localization. The MoI is related to the second moment of the mass distribution. Based on the idea, here we define 
$MoI(t) = \sum x^2 P(x,t)$, where $x$ labels the position in the lattice, $P(x,t)$ is the probability for the walker to be at $x$ at a given time $t$. Due to the mirror symmetry of the lattice and initial condition, the contributions from $x>0$ and $x<0$ are equal. 
The MoI is calculated at each step in the evolution after the rotation and translation operators have been applied.

\begin{figure}
    \centering
    \includegraphics[width=0.9\columnwidth]{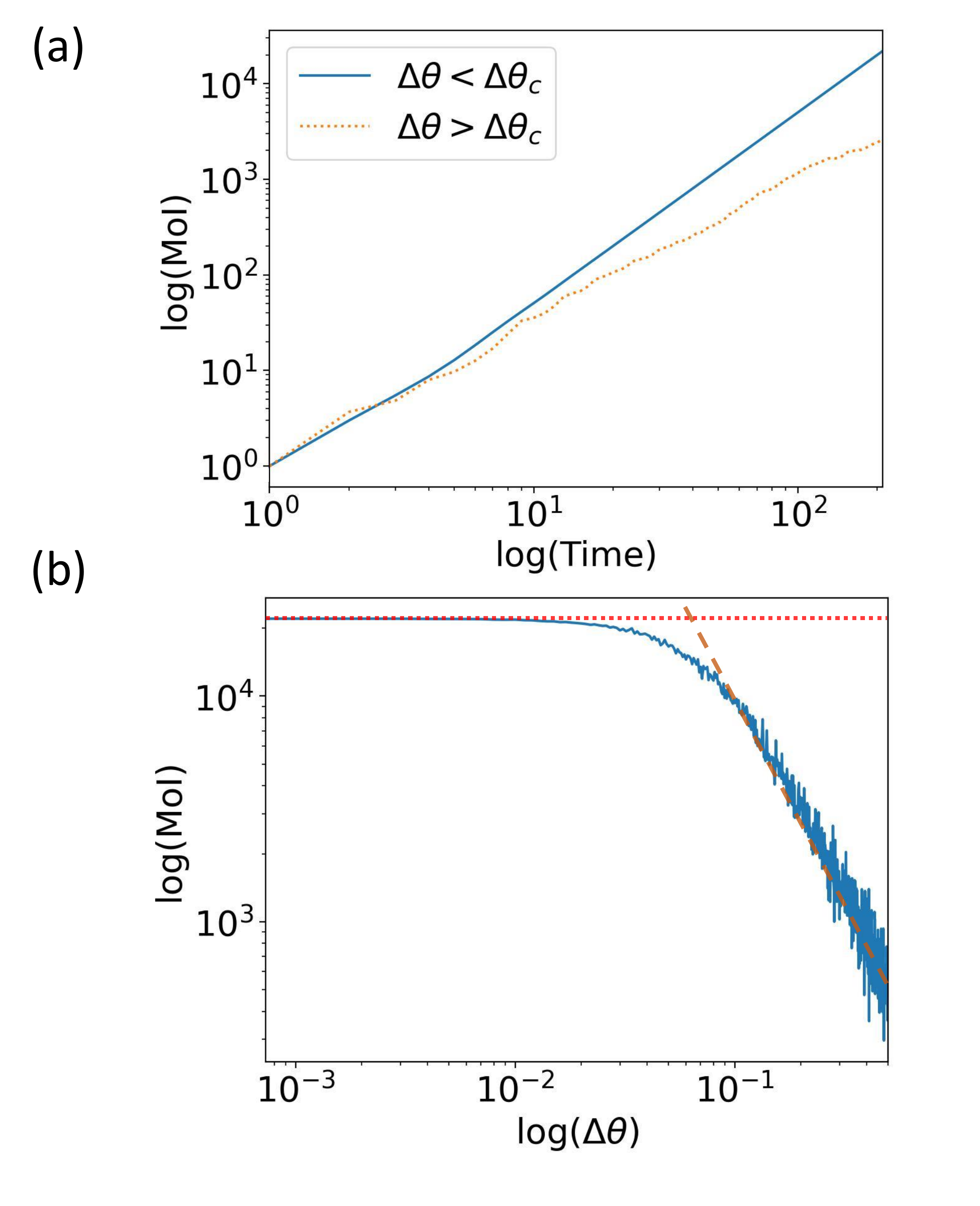}
    \caption{(a) Moment of Inertia (MoI) of a discrete random rotation QW with $\theta_{0} = \frac{\pi}{6}$ plotted as a function of time. The solid (dashed) line correspond to $\Delta\theta=0.001$ ($\Delta\theta=0.24$). (b) MoI of the final step at $N=210$ plotted against each value of $\Delta\theta$. The dashed lines indicate the two delocalized and localized regions of the walk.
    }
    \label{fig:MoIFigure}
\end{figure}

Fig.~\ref{fig:MoIFigure} (a) shows the MoI as a function of time for QW with discrete random rotation.  
In the long-time limit (say, $N>10$), the MoI of the case with small classical randomness is found to be proportional to the square of time ($N^2$). This agrees with the asymptotic behavior of the variance of a spreading quantum wavepacket~\cite{MQM} and similar analyses in the QW literature~\cite{STOC10.1145.380752.380757,PhysRevLett.92.120601}. As $\Delta\theta$ increases beyond a threshold, the long-time MoI starts to deviate from the $N^2$ dependence due to the localization of the probability distribution.

To extract the critical value, Fig. \ref{fig:MoIFigure} (b) shows the MoI at a fixed value of $N$ as a function of the random angle. 
Given the fixed $N$, if the randomness $\Delta\theta$ is smaller (or larger) than the critical value, the probability distribution spreads out (or localizes) in the delocalized (or localized) regime, and the MoI reaches (or falls below) the value proportional to $N^2$. 
The critical value $\Delta\theta_c$ can be found by locating the "kink" in the MoI curve. Similar procedures can be applied to QW with random continuous rotation or random translation.  Despite the quantitatively different behavior in the transition regime as shown in Fig.~\ref{fig:ProbFigure}, the MoI is able to indicate the critical value of $\Delta\theta$ or $P_r$ from the final probability distribution of the three cases studied here. However, the kink in the MoI sometimes may not be sharp, leaving room for possible deviation when determining the critical value.

\subsubsection{Inverse Participation Ratio}
\begin{figure}
    \centering
    \includegraphics[width=0.8\columnwidth]{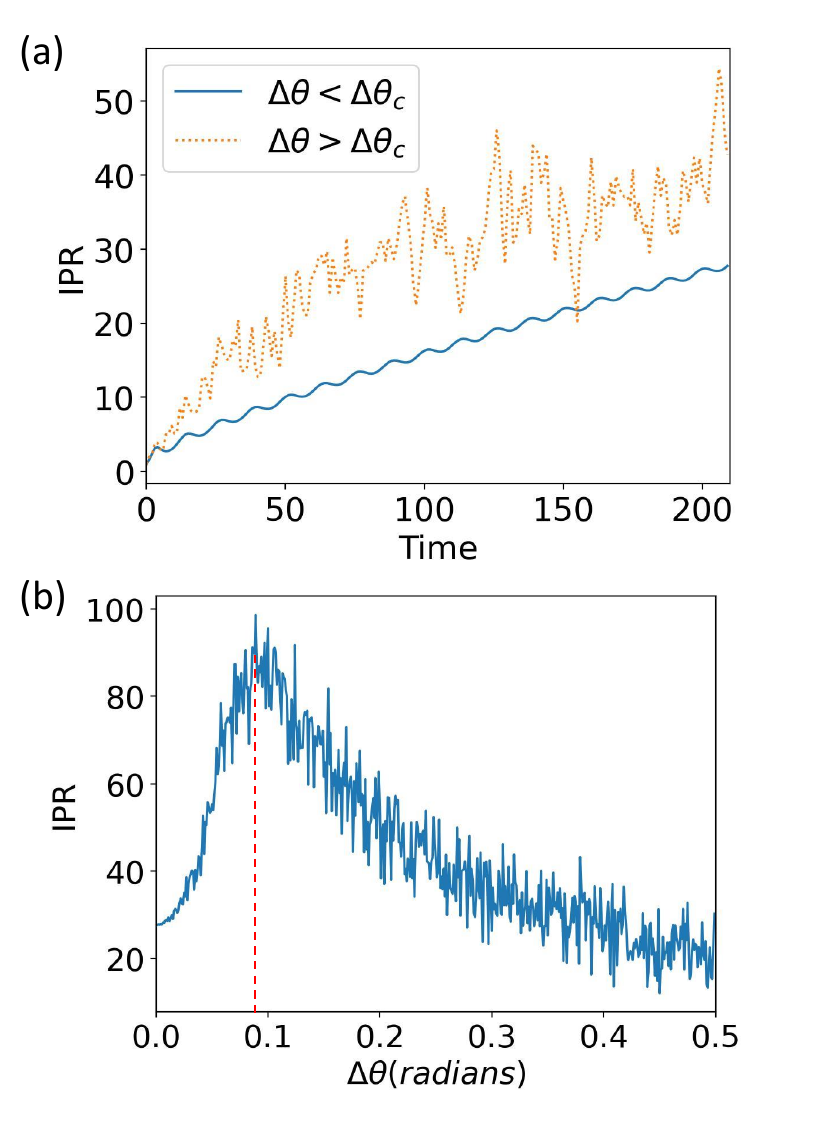}
    \caption{ Inverse participation ratio (IPR) of QW with discrete random rotation (a) as a function of time with $\theta_{0} = \frac{\pi}{6}$ and $\Delta\theta = 0.01$ (solid line) and $\Delta\theta = 0.23$ (dashed line) and (b) as a function of $\Delta\theta$ with $\theta_{0} = \frac{\pi}{6}$ and $N=210$. The vertical dashed line at the maximum offers an estimation of the transition.
    }
    \label{fig:IPRFigure}
\end{figure}

The IPR uses higher moments of the wavefunction to reveal the concentration of the distribution. We follow Ref.~\cite{1DFiberLoopALocal}  with the definition
\begin{equation}
IPR= \frac{(\sum_{x}||\psi_{x},+\rangle|^{2})^{2}}{\sum_{x}||\psi_{x},+\rangle|^{4}}.
\end{equation}
The IPR is maximal if the wavefunction spreads out evenly and decreases if there are spikes in the distribution. For QW with discrete random rotation, Figure~\ref{fig:IPRFigure} (a) shows the IPR curves as a function of time for selected $\Delta\theta$ below and above the critical value. As one can see, the IPR increases with time for both cases because the IPR identifies peak(s) in the distribution, regardless of one peak in the localized regime or two peaks in the delocalized regime.

Fig.~\ref{fig:IPRFigure} (b) shows the IPR for a fixed run time $\mathcal{N}_t$ as a function of the magnitude of classical randomness. 
The maximum of the IPR indicates the transition point, where the probability distribution is relatively flat across the space. However, one can see that the final-state IPR decreases for smaller or larger $\Delta\theta$. The reason is that the IPR does not discern if the wavefunction has one peak, as in the localized case, or two peaks, as in the delocalized case. Therefore, the IPR cannot differentiate the different phases. However, the coincidence of a relatively flat wavefunction at the transition allows us to use the maximum of the IPR as a quick check for the scaling of the critical behavior. For QW with random translation, the probability distribution in the transition region is relatively flat compared to the localized or delocalized regimes, despite its three-peak structure. Therefore, the IPR still shows a maximum in the transition regime and locates the critical value.

\subsection{Supervised machine learning}
Despite the success of the aforementioned methods for identifying the localization transition of quantum walk with classical randomness, one usually needs to examine a large number of data sets from the large parameter space for an estimation of the transition region, making the process very demanding on time and resources. Here we investigate the potential for machine learning to automatically identify and differentiate between the delocalization and localization behavior of quantum walk with various classical randomness.
Similar to the task of identifying phase transitions in physical systems by machine learning~\cite{PhysRevB.100.045129,PhysRevResearch.4.023005}, we implement two basic supervised learning methods to find the critical values of the quantum walk with classical randomness. We first feed to the machine the data from weak- and strong- randomness regimes with two different labels, respectively. After the machine can confidently differentiate the two cases according to the labels, we generate data with intermediate randomness and ask the machine for the probabilities of the new data belonging to the two groups. The critical point is determined by the maximal confusion point, where the machine outputs equal probabilities. 
The three supervised machine learning methods that will be tested are the support vector machine (SVM) multi-layer perceptron neural network (MLP NN), and the convolutional neural network (CNN). 

\subsubsection{SVM}

We begin with the SVM and train a supervised version from the sklearn linear SGD Classifier on two data sets with the same amount of run time and $\theta_{0}$. The first (or second) set of samples is from simulations with a small range of the classical randomness in the delocalized (or localized) regime supplemented with the label "0" (or "1").
The classifier is the "modified hubber" loss function to generate binary classification probabilities instead of the typical "hinge" loss function tailored towards pure true-false binary classification. When training the classifier, a fraction of the samples from each set were held back to act as a verification set that the trained classifier could correctly distinguish between the two final states. Once the classifier has been trained on the labeled data sets, we preform a test using the reserved samples and checking the classification probabilities to confirm that it is able to distinguish between the delocalization and localization behavior. The small range for generating the training data in the respective regimes with weak and strong randomness has been adjusted to reach high accuracy in the training.
We remark that if the range is too small, the data may not give the machine enough information about the configurations. However, if the range is too large, the information of the transition regime may be involved in the data and affect the reliability of the decision of the critical value.

\begin{figure}[t]
    \includegraphics[width=\columnwidth]{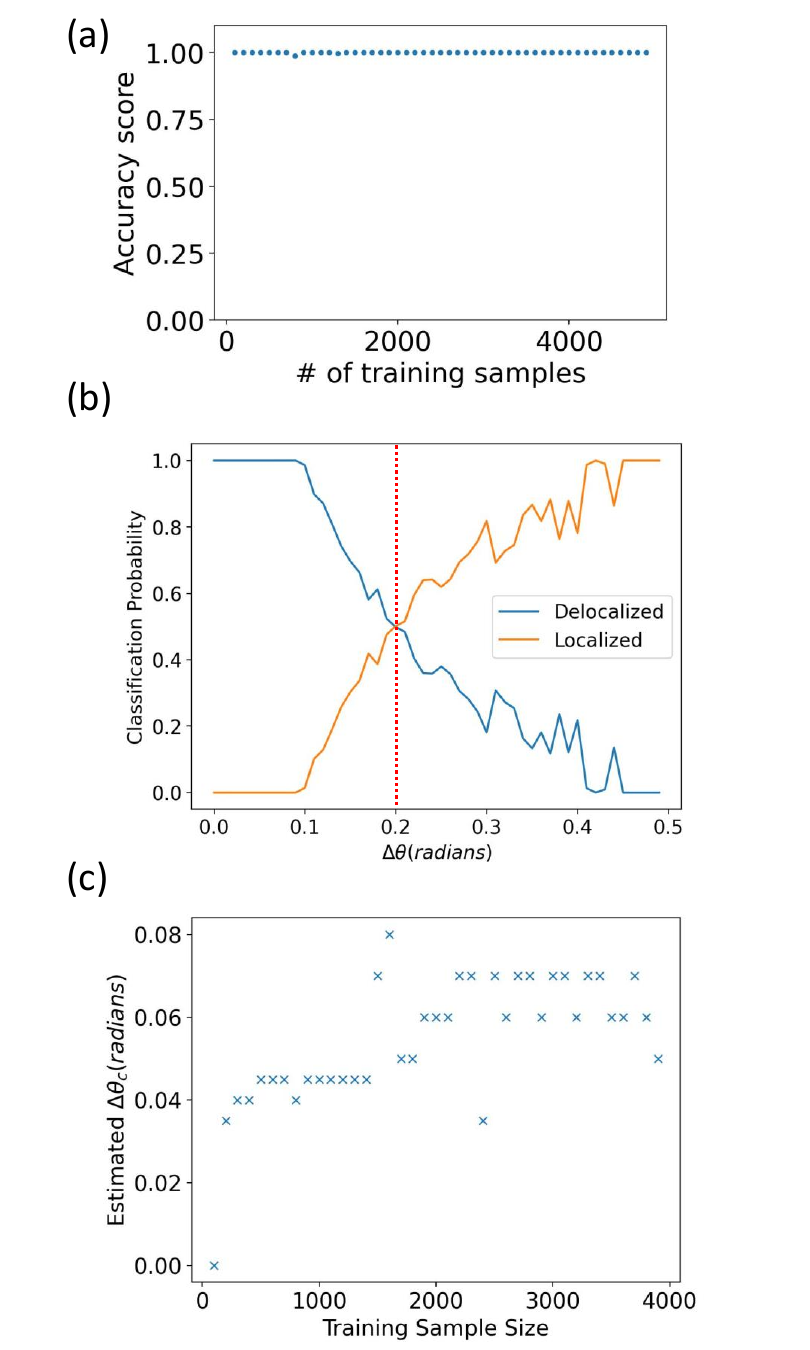}
    \caption{(a) For the split training and testing data, the SVM's  accuracy dependence on the training sample size is relatively flat. (b) Testing the SVM to classify the localization transition of QW with discrete random rotation. Here $\theta_{0} = \frac{\pi}{6}$ and $N = 100$. The vertical dashed line indicates the maximal confusion, which locates the value of $\Delta\theta_c$. (c) Dependence of the critical value determined by the SVM on the size of the training data set. 
    }
    \label{fig:MLProb-Samplesize}
\end{figure}

During the training of the ML model, the full set of training data was split with an 80/20 ratio between the training set and a testing set to check the model's accuracy. Once the model was trained by the training data set, we checked its accuracy in predicting the labels on the reserved, unseen probability distribution of the testing set and its potential dependence on the amount of samples used within the training data set. The accuracy of labeling the unseen testing set is shown in Fig. ~\ref{fig:MLProb-Samplesize} (a), which shows a near 
perfect prediction accuracy already with a small sample size. Given that the model is trained by two fairly distinct (one-peak vs. two-peak) distributions, this nearly perfect accuracy is not surprising and only confirms that our model is able to distinguish unambiguously between the localized and delocalized regions.


After the training and testing, the additional data set used to identify the transition region is generated throughout the parameter regime with intermediate randomness. At each $\Delta \theta$ or $P_{r}$, the probability distribution is obtained up to time step $N$. 
Using a classifier trained on a given $N$, the additional data in the transition regime are fed into the classifier. 
After plotting the probabilities associated with the data in the intermediate regime as shown in Fig. \ref{fig:MLProb-Samplesize}, the critical value ($\Delta\theta_{c}$ or $P_{r,c}$) near the transition point is taken to be the maximally confusing point, where the two classification probabilities are equal.

We have also checked the influence of the amount of samples used for training the classifier on the predicted critical value. To find a minimum sample size, we were mainly concerned with the relative stability of the estimates of the critical values from the classifier. As seen in Fig.~\ref{fig:MLProb-Samplesize} (c), a relatively small training data set is already able to distinguish the overall trend of the localization transition, but the results are more convergent as the total amount of samples used increases. 
For the localization problem, we found that with a training set of about 2000 samples, the estimates for the critical value already settle into a more stable pattern with few extreme estimates. Increasing the size of the training data further does not lead to visible improvement.
In the following, we will present results with a training data set of sample size 1800, as it achieves the minimization of the outlying estimates while being computationally manageable for a large range of $\Delta \theta, P_{r}$ values.

\subsubsection{Neural networks - MLP}
Next, we implement a fully connected feed-forward neural network called the multi-layer perceptron neural network (MLP NN) as a supervised learning method to identify the localization transition of QW with classical randomness. We use the network provided by sklearn's MLPClassifier which implements a multi-layer perceptron binary classifier network. In constructing this classifier, we follow the same procedure that was outlined and used with the SVM method. Since we wish to compare with the SVM, we train and employ the NN on the same data sets that were used with the SVM. When training the NN, sklearn's GridSearchCV library was used to help in working a range of given hyper-parameters to find potential combinations that lead to either greater efficiency in the computational resources needed for convergence, or ones that would lead to greater accuracy in the critical value estimates. Although there is a broad range of hyper-parameter available, we focused on two of the more immediate parameters, the hidden layer sizes and the regularization parameter $\alpha$. 

In our investigation, we used a network that consisted of six total layers: One input layer whose number of neurons is automatically adjusted to the size of the input data, four hidden layers, and one output layer that contains two neurons for the binary classification task. The four hidden layers were taken to have 400, 200, 100, and 50 neurons respectively for the simulation data with time up to $N=1000$, as we found this to be a relatively simple "middle ground" that did not add a great amount of computational time while improving the overall estimates. As we increased the simulation time from $N=80$ to $N=1000$, we found that the network was able to make reasonable estimates with these hidden layer sizes and that adjusting the any of the layer's neuron number to the increasing lattice size showed a negligible improvement on the final estimates. Moreover, since the first layer of the network 
is automatically shaped to accommodate the size of the training data, there was no processing needed before the data sets could be used in the network other than the normalization mentioned below. After some trial and error, the regularization parameter was taken to be $\alpha = 0.001$ since this combination of parameters provided reasonable accuracy.
While both the SVM and MLP NN methods take the critical value to be the point of maximal classifier confusion, the NN was found to show a more sudden change in the classification probabilities rather than the more gradual transition found through the SVM. Thus, for the MLP NN, we take the critical value to be the first point where the classification probability of being delocalized falls below $50\%$ as an estimation of the maximally confused point.

\subsubsection{Neural networks - Convolutional}
Lastly, we implemented a supervised convolutional neural network using the TensorFlow framework and libraries to estimate the localization transition by using the same training and testing data as those used in the SVM and MLP NN. Here the CNN was created to provide binary classification probabilities and consists of four 1-dimensional, convolutional layers with a final dense layer that uses two neurons and the "softmax" activation function to provide the binary classification probabilities. In order to achieve binary classifications for the localization transition, we chose to use the "Sparse Categorical Crossentropy" loss function as it is the recommended loss function for this type of classification task. 

The number of neurons used in the convolutional layers was chosen to be a function of the number of the lattice size with the first layer having $2N + 1$ neurons that matches the lattice size and the second, third, and fourth layers having half, one-fourth, and one-eighth the initial number of neurons rounding up to the nearest integer, respectively. After each layer, we also used TensorFlow's "Dropout" regularization function to randomly reset the output of multiple neurons in an effort to reduce the chance of overfitting in the model. In practice this acts to reduce the number of potential parameters created in the network as it is being trained. This was preformed three times with the first two dropping $50\%$ of the neurons, and the final operation dropping only $20\%$ of the final layers neurons. In practice, this layer structure will automatically adapt to the increasing number of lattice sites in the simulations. Since the layers automatically adjust themselves to the lattice size of the given data set, we did not preform any processing on the data other than the normalization to meet the machine-learning convention mentioned below. While the model was re-trained for each increase in the lattice size to adjust the number of neurons in each layer, no changes were made to the overall network structure for data from larger lattice sizes as we found no tangible improvement in the estimates as the number of nodes increased.

For QW with random translation, we found it necessary to add L2 kernel regularization to each layer in the CNN to reduce the network's tendency to overfit these more varied probability distributions. Similar to dropout regularization, L2 (or ridge) regularization attempts to limit the number of potential parameters during training to reduce the chance of overfitting. However, unlike the dropout method, L2 regularization is preformed on the network's loss function and penalizes the model when it begins creating a large number of parameters. Within our work we used the packages provided by TensorFlow and a regularization factor of $\lambda = 0.01$ was used. Using this change, we note that the regularization improved the estimations for smaller simulation times ($N=80 $ to $N=500$), where there is a greater range of $P_{r}$ values in the transition regime, and thus greater variability in the probability distributions when locating the transition point.

Since the size of the lattices of QW considered in this work was always an odd number, we chose the size of the convolution kernel to be three as this removed the need for the data to be padded in order to make sure the entirety of the probability distribution was used in the convolution. For each new feature set, the model was trained for 10 epochs as we found this to be a satisfactory amount of epochs that reduced the training time while ensuring the model was well trained. The rest of the training and testing procedures follow the same procedure that was used with the two previous machine learning methods. The estimation of the critical point is similarly found through the same way as the MLP NN method.

\begin{figure}
    \centering
    \includegraphics[width=0.9\linewidth]{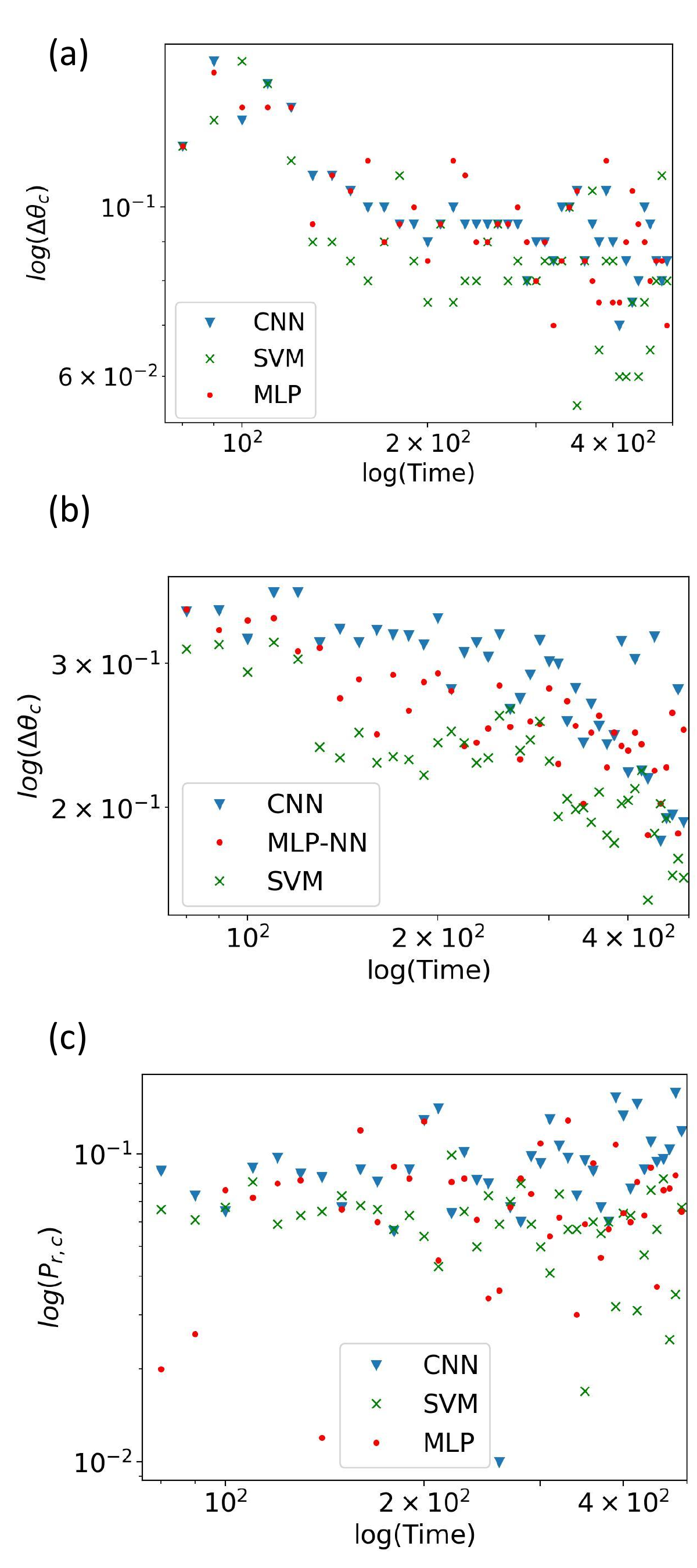}
    \caption{Critical values of the localization transition from the SVM (crosses), multi-layer perceptron NN (dots), and the convolutional NN (triangles) for QW with (a) discrete random rotation, (b) continuous random rotation, and (c) random translation.  All panels are with $\theta_{0} = \frac{\pi}{6}$.
    }
    \label{fig:SVM-NNFig}
\end{figure}

\subsubsection{Comparison of machine learning methods}
We estimated the critical values for each of the three types of QW with classical randomness (discrete random angle, continuous random angle, and random translation) using the same training and testing data sets. A comparison between the three learning methods will demonstrate their abilities to generate reliable estimates and more importantly pick up on the scaling trend from the localization transition. Additionally, we investigated the normalization of the data for each method. The probability distribution $P(x)$ of the QW has the default normalization that $\sum_{x}P(x)=1$. However, the convention in supervised machine learning is to normalize the distribution so that $P(x)$ spans the range of $[0,1]$. Therefore, we normalize each probability distribution by its maximal value to ensure $P_{max}=1$ in each data set to match the normalization that is used within areas like image recognition where machine learning is heavily used.
The normalization was preformed on both the training and estimation data sets. We found that while the normalization did not improve the estimates given by the SVM, the estimates from the MLP NN and CNN did show much less overall variation because neural networks are usually designed with the magnitude normalization in mind. 

The critical values shown in Fig. \ref{fig:SVM-NNFig} confirm that the three supervised machine learning methods are capable of classifying the localization transition. Moreover, their estimations of the critical values are fairly close. Interestingly, a more complex neural network design like the CNN produces results that are similar to the simpler SVM and MLP NN methods.
On the other hand, each method appears to produce the same effect in their estimates in that rather than picking out a singular point like the manual methods of the probability distribution, MoI, and IPR, the machine learning programs tend to generate their estimate from a small range around the transition point. Later on, we will see this is likely the cause of the lower  exponents of the localization reported from the machine learning methods. 

\section{Comparison and discussion}\label{sec:Comparison}

\subsection{Comparison of different methods}
\begin{figure}[t]
    \centering
    \includegraphics[width=0.87\columnwidth]{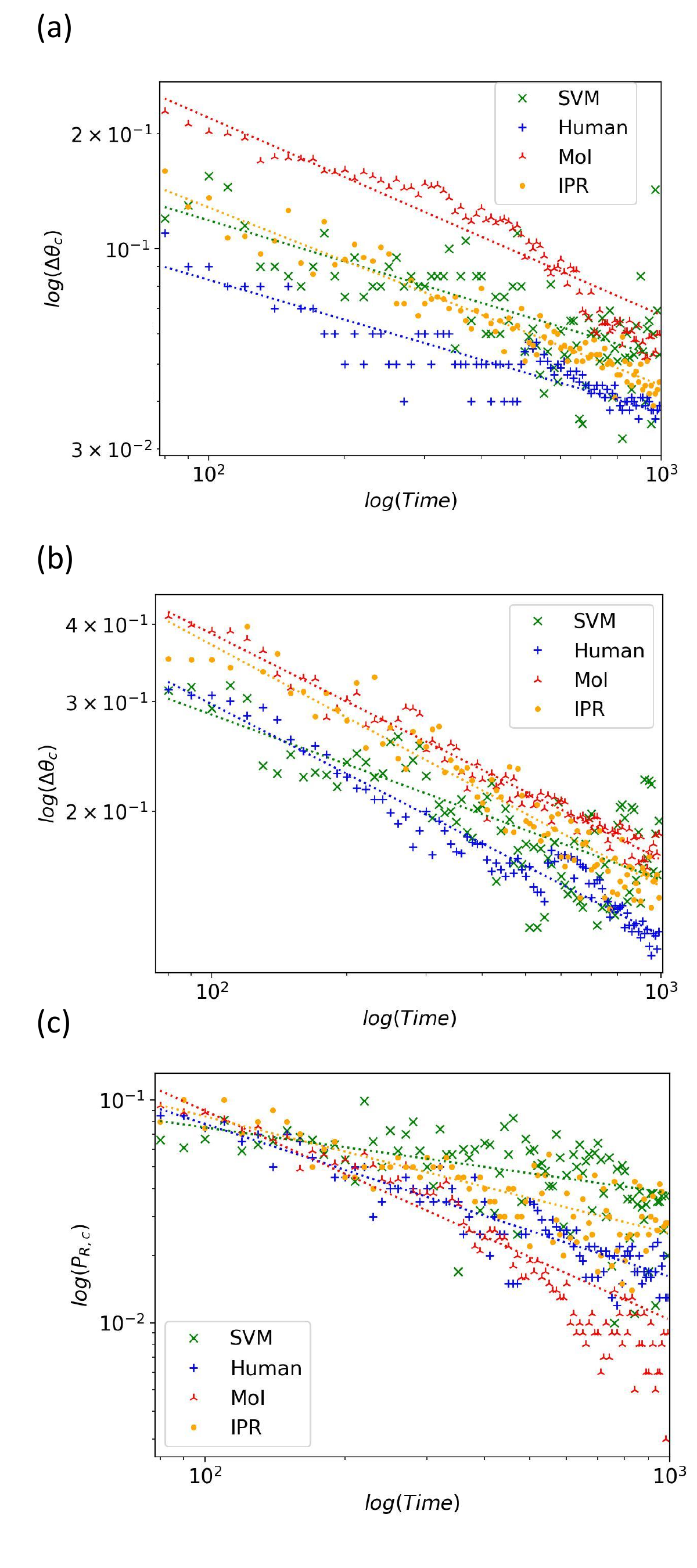}
    \caption{Comparison of the critical values separating the localized and delocalized regimes estimated from the four methods for QW with (a) discrete random rotation, (b) continuous random rotation, and (c) random translation. The dashed lines show the power-law fit. Here "Human", "MoI", "IPR", and "SVM" correspond to the probability distribution, momentum of inertia, inverse participation ratio, and SVM, respectively.
    }
    \label{fig:Multi-methodResults}
\end{figure}

In our simulation, the evolution time determines the size of the lattice that the quantum walker can explore. To determine the scaling behavior of the localization, we first pick a fixed evolution time $N$ and run the simulations for a range of $\Delta\theta$ or $\Delta\theta_M$ for the random-rotation case or $P_r$ for the random-translation case. By examining manually the probability distribution, MoI, or IPR, we locate where the localization transition occurs and label the critical value $\Delta\theta_c$ or $P_{r,c}$ for the chosen $N$. Next, we plot the critical $\Delta\theta_c$ or $P_{r,c}$ versus $N$ of the three manual methods and SVM in Fig.~\ref{fig:Multi-methodResults}.  
By fitting each set of critical values with power laws,
the exponents of localization are extracted.
Table~\ref{Table-exp} summarizes the exponents from the six methods for the three types of randomness. The relatively small uncertainties of the exponents only reflect that the squared residuals of the fitting from the scattered critical values as shown in Fig.~\ref{fig:Multi-methodResults} cannot be reduced further by adjusting the slope on the log-log plot. We have checked that adding more points from the analysis in the small (or large) parameter regime or increasing the simulation time from $N=500$ to $N=1000$ does not lead to qualitative differences in the scaling behavior and exponents. 
Moreover, we have checked that the value of $\Delta\theta_c$ is insensitive to the value of $\theta_0$ for QW with random rotation, as long as $\theta_0$ is away from $0$ or multiples of $\pi/2$.

Fig.~\ref{fig:Multi-methodResults} (a) shows the critical values of $\Delta\theta_c$ of QW with discrete random rotation. The three manual methods from the final probability distribution, MoI, and IPR find a similar trend in the critical values, indicating a power-law dependence of the localization with the randomness magnitude. 
A closer examination shows that the exponent from the SVM is comparable to those from the manual method based on the final probability distribution and the IPR. However, the SVM result exhibits large fluctuations of the critical values as the simulation time increases. This is likely due to the uncertainty in determining where the structures of the probability distribution change in large-size data. Meanwhile, the MoI produced a higher exponent compared to all others, possibly due to the less-sharp region in determining the critical value mentioned in the discussion of Fig.~\ref{fig:MoIFigure} (b).
Since the network-based machine-learning methods and SVM produce similar results as shown in Fig.~\ref{fig:SVM-NNFig}, we do not show the NN results again to avoid  overcrowding the panel.

Next, Fig.~\ref{fig:Multi-methodResults} (b) shows the critical values  of $\Delta\theta_c$ from the three manual methods and the SVM for QW with continuous random rotation.
Despite the difference in the classical distribution (discrete vs. continuous), the resemblance between Fig.~\ref{fig:Multi-methodResults} (a) and (b) indicates that similar scaling behavior is likely behind the localization. Some step-like patterns are due to the limited resolution of the parameters, and we have checked that increasing the resolution does not lead to significant changes of the results. For this case, the exponents from the three manual methods are close to each other while those from the three machine-learning methods are close to each other. The exponents from the SVM and MLP NN fall below the other methods while that of the IPR is slightly above.

Finally, Fig.~\ref{fig:Multi-methodResults} (c) shows the critical values of $P_{r,c}$ of QW with random translation from the manual methods and SVM. 
The analysis hints that the scaling behavior of the localization transition may be general. However, there is a much greater amount of variation across 
different methods. 
As shown in Table~\ref{Table-exp}, the MoI agrees with the IPR while the method based on final distribution produced a higher value. On the other hand, the SVM produced a lower estimation of the exponent while the two network-based methods exhibit systematical deviation and could not catch the trend of the exponent. The substantial variation is due to an increase in the fluctuations within the probability distributions of QW with random translation when compared to those from QW with random rotation. Nevertheless, the power-law dependence of the data supports the general idea of scaling behavior behind the localization transition. 

The SVM only produces exponent comparable to the manual methods for QW with discrete random rotation and underestimates it in the other two cases. 
On the other hand, the two NN-based methods show exponents close to the SVM for the two QW cases with random rotation, but their exponents deviate from the other methods for QW with random translation because for small values of $N$, the NN-based methods tend to generate estimates that are slightly into the delocalized regime but give estimates closer to the localized regimes for larger $N$ values, possibly because of the noisy behavior in the probability distribution.
The systematic under- and over- estimations flatten out the scaling trend and render the systematically lower scaling exponents. As shown in Table~\ref{Table-exp}, the manual methods of QW with random translation also tend to show larger deviations compared to the results of QW with random rotation, suggesting QW with random translation could be a challenging problem for classification methods.

\begin{table}
\begin{tabular}{|c|c|c|c|}
    \hline
    & $\Delta\theta_{c}$ (discrete)  & $\Delta\theta_c$ (continuous) & $P_{r,c}$ \\ \hline
    SVM  & $0.36 \pm 0.04$ & $0.25 \pm 0.02$ & $0.28 \pm 0.04$ \\ \hline 
    MLP NN & $0.32 \pm 0.02$  & $0.25 \pm 0.01$  & $0.07 \pm 0.06$ \\ \hline
    
    CNN & $0.39 \pm 0.07$ & $0.32 \pm 0.04 $ & $-0.08 \pm 0.06$  \\ \hline

    MoI  & $0.62 \pm 0.02 $ & $0.36 \pm 0.02$  & $0.51 \pm 0.04$\\ \hline

    Human & $0.32 \pm 0.01$ & $0.36 \pm 0.01$ & $0.69 \pm 0.03$ \\ \hline

    IPR  & $0.46 \pm 0.02$ & $0.41 \pm 0.01$ & $0.52 \pm 0.02$ \\ \hline

\end{tabular}
\caption{\label{Table-exp} 
Exponents of the power-law dependence of the critical value on the size of QW. The "Human" method is from the manual determination of the probability distribution.
}
\end{table}

As shown in Fig.~\ref{fig:ProbFigure}, the transition regime of QW with random translation is complicated due to the coexistence of the central peak and the two side peaks.
In an effort to dissuade the MLP-NN and CNN from systematically under- or over- estimating the transition points, we split the probability distributions up into three regions: (1) $-N<x<-N/2$, (2) $-N/2<x<N/2$, and (3) $N/2<x<N$. Regions (1) and (3) contain the left- and right-most peaks seen in panel (e) of Fig.~\ref{fig:ProbFigure} while region (2) contained the neighborhood of the origin showing the evolution from a flat, delocalized distribution to the localized central peak. The three regions were then separately used as the new training data sets to generate the new transition region estimates. Using these sets of disseminated data for the two NN-based methods, we found no observable improvements in the transition estimates using either of these regions. The results also help us see that the two NN-based methods used here weight in both local and global features of the data rather than catching some local prominent variations. 
With the NN-based methods, however, this lack of improvement further highlights the difficulty that the commonly available machine-learning methods may face when attempting to differentiate systems that have mixed quantum and classical probabilities.

\subsection{Similarities and differences with Anderson localization}
The Anderson localization~\cite{PhysRev.109.1492} is a textbook example of localization induced by classical randomness in quantum systems. It is usually described by the random hopping model on a lattice with the Hamiltonian
\begin{eqnarray}
H_A=\sum_{\langle i,j\rangle}t_{ij}c^{\dagger}_i c_{j}+ H.c. + \sum_{j}V_j c^{\dagger}_j c_j.
\end{eqnarray}
Here $c^{\dagger}_j$ ($c_j$) is the creation (annihilation) operator on site $j$, $t_{ij}$ is the hopping coefficient between the nearest-neighbor pair $\langle i,j\rangle$, and $V_j$ is the onsite potential. The classical randomness of the hopping model can be introduced by drawing $t_{ij}$ or $V_{j}$ from uniform distributions within the range $(-W,W)$ in each step. While the 1D and 2D ground states in the thermodynamic limit are localized~\cite{AnderLocalBook}, there can be delocalized states in 3D above a threshold energy.
The classical randomness of the random hopping model is quenched in real space as the system evolves. In contrast, the classical randomness of QW discussed here is uniform in space but varies with time as the system evolves. Despite the different ways of introducing classical randomness, localization transitions are observed in both cases.
We mention that Ref.~\cite{Carrasquilla2017} showed that machine learning can differentiate the phases across the Anderson localization in a quasi-periodic lattice, where the deterministic parameters induces the localization transition, in contrast to the typical Anderson localization that mixes quantum and classical probabilities.

Ref.~\cite{LocalizationMatterWave} provides a functional form for the 1D localization length of the Anderson model. A relation between the localization length $L_{loc}$ and the amplitude of the disorder $W$ is written in the form $L_{loc} \propto W^{-2} \Rightarrow  W \propto L_{loc}^{-0.5} $. Their experimental results support this scaling behavior. In our results of QW with classical randomness, we also identified a localization transition and summarized the the exponents extracted from the localization transition in Table~\ref{Table-exp}. We remark that Ref.~\cite{PhysRevB.84.195139} shows that if certain symmetry is present in QW, it may escape the localization from spatial disorder but remains localized with temporal disorder. 
Moreover, Ref.~\cite{PhysRevLett.106.180403} realized photonic QW up to 28 walk steps and contrasted static and time-dependent phase disorders. Their results confirmed the different probability distributions due to the different types of randomness, demonstrating the subtle difference between static and dynamic disorders. Therefore, a more unified view of spatial and temporal randomness in quantum systems still awaits future research.


\subsection{Implications}
Our analysis of QW with classical randomness has shown possible scaling behavior of the localization transition, which transcends different methods for inducing classical randomness. 
Realizations and measurements of quantum walk with classical randomness have been achieved. For example, Ref. ~\cite{PhysRevLett.121.070402} used atomic Bose-Einstein Condensate (BEC)  to experimentally implement a discrete-time QW. 
Classical randomness called "noise" has been introduced through the phase factor of the coin operator, resulting in a change of the momentum distribution associated with the transition from quantum to classical walk dynamics.  
In Ref.~\cite{PhysRevLett.104.100503}, a single
ion in a linear Paul trap 
allows for the control of its internal states through a series of controlled $\frac{\pi}{2}$ pulses. 
Like Refs.~\cite{1DFiberLoopALocal, PhysRevLett.121.070402}, they introduce classical randomness through the coin phase parameter, using a random noise generator  for each of the  pulses, and show the change from the quadratic propagation speed of QW to the linear speed of a classical random walk.
In Ref.~\cite{PhysRevLett.104.153602}, optical implementations of a discrete-time QW using entangled pairs of photons split at each step with one being directly used in the walk while the other acting as a trigger for the measurements. 
By differing the angle of the mirrors in each step, the photons can decohere. Over a small number of walk steps ($N = 6$), the probability distributions are shown for QW and for the decohered walk.


The under-estimate of the exponent of QW with random translation from the MLP NN and CNN provides an example that mixing classical and quantum probabilities is still a challenging task for machine learning methods. While we can manually resort to physical quantities such as the MoI or IPR to pinpoint the critical point, the machine associates certain features with the labels for classification. We have seen the classification may be constrained by limitation of the data size or distraction from the fluctuating part in the bulk of the data. Our results thus present an open challenge for machine learning to better differentiate systems with both classical and quantum probabilities.

In our investigations with neural-network based classification, we had tested the MLP NN classifier as it provides a plain and immediate interface to  various problems. While this classifier is able to generate classification predictions of a certain level of accuracy, it may show a tendency to overfit the predictions ~\cite{IEEE.2000.857823, NNOverfit97, IEEE.4696782} and thus bias the final results. We also tested the CNN based classifier \cite{IEEE.018.8540626, IEEE.72.554195, SPIE.10.1117.12.2262589}, which has previously shown improvement over the MLP classifier in the overall accuracy of estimations and reduced potential for overfitting when used on more complex data sets \cite{IEEE.2018.8484751, AAAI.10.5555.2283516.2283603, park17c_interspeech}. Interestingly, the CNN produced similar results close to the MLP NN when identifying the critical values of the QW with classical randomness studied here.
Therefore, benchmarking machine learning with the localization transition is a problem worth future investigations.

\section{Conclusion}\label{sec:Conclusion}
The localization  of QW with classical randomness allows us to explore the scaling behavior of the transition. While the localization transition realizable in many experiments can be characterized manually by physically inspired quantities, such as the MoI or IPR, the supervised learning methods using the simple SVM and MLP NN and the more sophisticated CNN also catch the transition from the simulation data. 
Depending on the type of randomness, 
the machine-learning methods may agree or underestimate the exponents compared to the manual methods. A challenge emerges as the two NN-based methods seem to deviate systematically for the case of QW with random translation, possibly due to the more structured probability distributions in the transition regime. Our results may inspire future research on  systems with mixed quantum and classical probabilities and applications of machine learning to those systems.

\begin{acknowledgments}
We thank David Colby for some early explorations of the problem. C. C. C. thanks the KITP at UCSB for its hospitality. This work was partially supported by NSF No. 2310656.
\end{acknowledgments}

\bibliographystyle{apsrev}

\end{document}